\documentclass[12pt]{article}
\usepackage{graphicx,amsmath}
\usepackage{units}

\newlength{\dinwidth}
\newlength{\dinmargin}
\renewcommand{\vec}[1]{\boldsymbol{#1}}

\def\lapproxeq{\lower .7ex\hbox{$\;\stackrel{\textstyle                                                    
<}{\sim}\;$}}                                                    
\def\gapproxeq{\lower .7ex\hbox{$\;\stackrel{\textstyle                                                    
>}{\sim}\;$}}                                                    
\def\be{\begin{equation}}                                                    
\def\ee{\end{equation}}                                                    
\def\bea{\begin{eqnarray}}                                                    
\def\eea{\end{eqnarray}}

\def\GeV{\rm GeV}

\def\sh{\hat s}
\def\sh2{{\hat s}^2}

\begin{document}
                                                    
\titlepage                                                    
\begin{flushright}                                                    
IPPP/19/15  \\                                                    
\today \\                                                    
\end{flushright} 
\vspace*{0.5cm}
\begin{center}                                                    
{\Large \bf Exclusive vector meson production}\\
\vspace{0.5cm}
{\Large \bf  in heavy ion collisions}\\
\vspace*{1cm}
                                                   
V.A. Khoze$^{a,b}$, A.D. Martin$^a$ and M.G. Ryskin$^{a,b}$ \\                                                    
                                                   
\vspace*{0.5cm}                                                    
$^a$ Institute for Particle Physics Phenomenology, University of Durham, Durham, DH1 3LE \\                                                   
$^b$ Petersburg Nuclear Physics Institute, NRC Kurchatov Institute, Gatchina, St.~Petersburg, 188300, Russia

\vspace*{1cm}                                                    
                                                    
\begin{abstract}                                                    
                                                 
We discuss the salient features of exclusive vector meson production in heavy ion collisions at LHC energies. Special attention is paid to the space-time picture of the process.
We account for both coherent and incoherent contributions. The explicit quantitative predictions are
given for the $\rho$-meson differential cross section in lead-lead collisions in different kinematical configurations relevant for the LHCb and ALICE experiments.

\end{abstract}

\vspace*{0.5cm}                                                    
                                                    
\end{center}

 \section{Introduction}
 
\begin{figure} [t]
\begin{center}
\includegraphics[trim=2.0cm 6cm 2cm 4cm,scale=0.5]{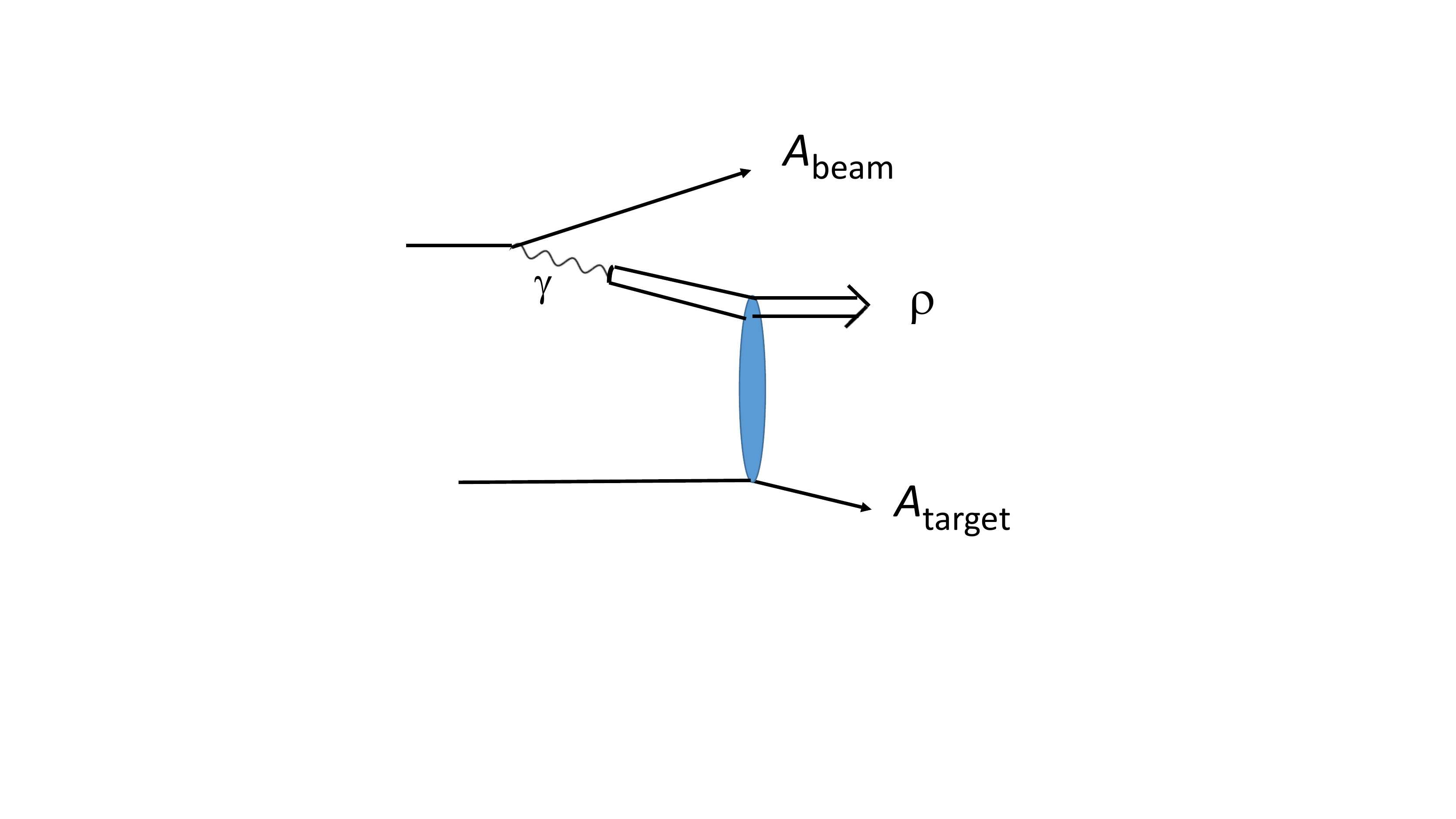}
\caption{\sf Exclusive $\rho$ production in heavy ion collisions, $AA\to A+\rho +A$, is the dominant contribution.}
\label{fig:1}
\end{center}
\end{figure}

An attractive feature of ultraperipheral vector $\rho$  meson production in heavy ion high-energy collisions is that the dominant contribution comes from the {\it purely exclusive} channel
\be
AA ~\to~A + \rho + A~.
\ee
Here $A$ is the heavy ion and the + signs denote the presence of large rapidity gaps. This observation was emphasized in ref.~\cite{KN} and studied experimentally by the STAR collaboration at RHIC
\cite{Adler:2002sc}~-~\cite{Adamczyk:2017wyc}
and by the ALICE collaboration at the LHC
\cite{ALICE}.  
Recall that 
 $\rho$ meson production can essentially be described using the Vector Dominance Model (VDM)~(see e.g. \cite{Sakurai:1960ju,Bauer:1977iq})
 as a $\gamma\to \rho$ transition followed by elastic $\rho$ scattering with the `target' $A$ ion. The process is pictured in Fig.~\ref{fig:1}.
  Note that here we deal with (in some sense) a {\it strong long range interaction}. Indeed, the factor $Z$ almost compensates the smallness of the QED coupling, and in terms of VDM we have the strong interaction of $\rho$ (or $\omega,\ \phi,...$) mesons which thanks to the photon propagator can take place at large distances both in coordinate (impact parameter $b$) space and in rapidities.
  
  The process is mainly exclusive. That is the probability of incoming ion dissociation is suppressed. For the ion which emits a photon the probability of an `inelastic' radiation is proportional to $Z$. It is factor of $Z$ smaller than the probability of coherent emission which is proportional to $Z^2$. On the other hand
  in the interaction with another (`target') ion
 the elastic $\rho A$ cross section is close to the full area  
of the heavy ion, $\sigma_{\rm el}\simeq \pi R^2_A$, whereas the cross section for dissociation comes only from the peripheral ring around the ion $\sigma_{\rm diss}\simeq 2\pi R_Ad$, where $d$ is the width of the ring. That is we expect a factor $2d/R_A$ suppression.

  For these reasons the exclusive contribution to $\rho$ meson production, $AA\to A+\rho+A$, will dominate. As a consequence 
  the experimental measurement of the process $AA\to A\rho A$ will allow 
 the observation of diffractive structure in the differential cross section, $d\sigma (\rho A)/dt$, for $\rho A$ {\it elastic} scattering.  
 
 In the present paper we describe the space-time picture of the process.
 We discuss the incoherent background caused by the elastic elementary
  $\rho+n\to\rho+n$ interactions and that caused by the possibility of nucleon 
  $n\to n^*$ dissociation; i.e. $\rho+n\to\rho+n^*$.
  Finally we show the cross sections expected at the LHC for few different kinematics.
  
\begin{figure} [h]
\begin{center}
\includegraphics[trim=1cm 5cm 0cm 5cm,scale=0.5]{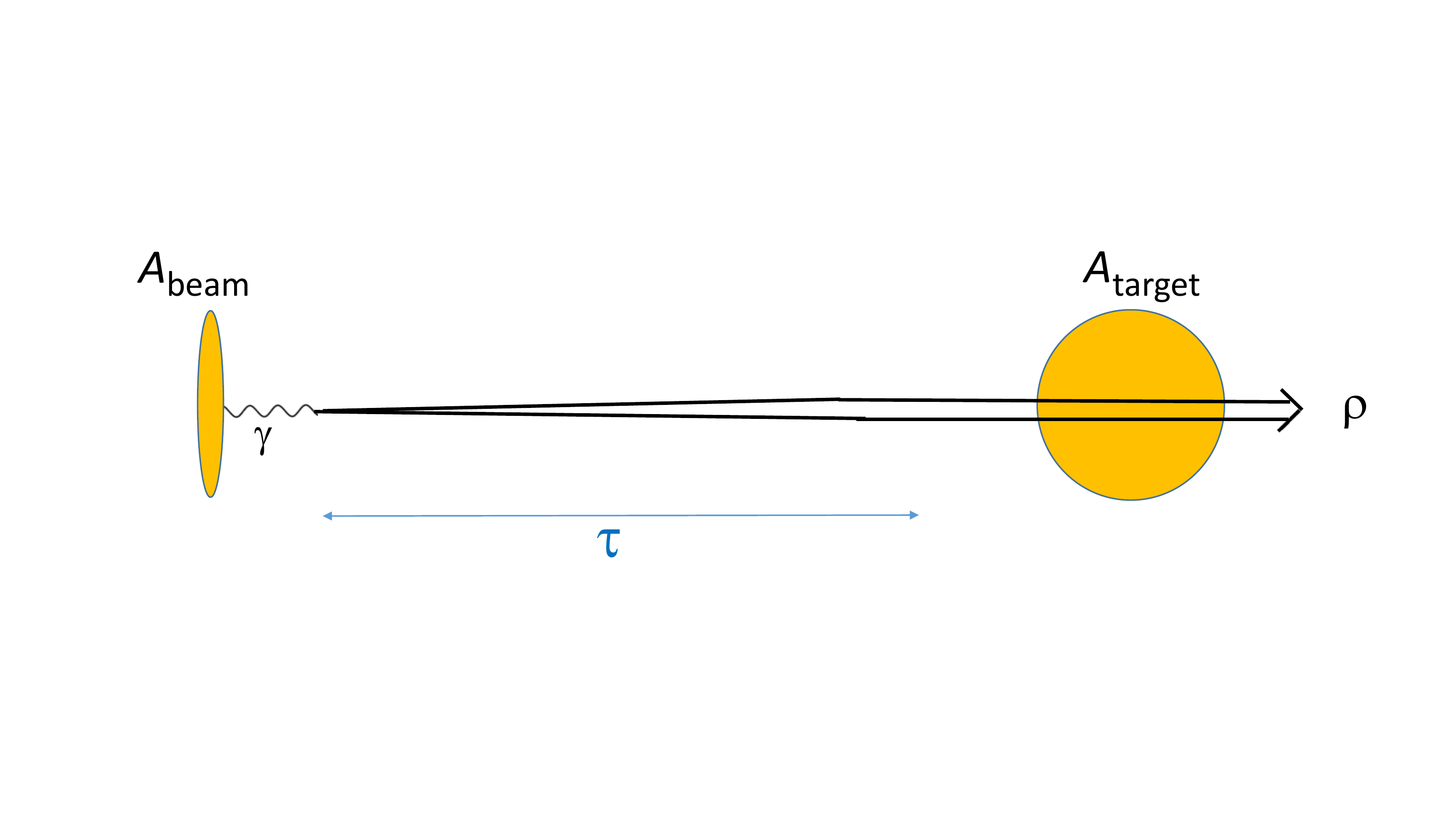}
\caption{\sf A sketch of the exclusive $\rho$ production process in the rest frame of the target heavy ion. The $q\bar{q}$ pair is only able to form the $\rho$ meson when its transverse separation becomes large enough ($\sim 0.5$ fm). At high energies this results in a long formation time $\tau$ for the $\rho$ meson.  }
\label{fig:2}
\vspace*{-0cm}
\end{center}
\end{figure}

The $\rho$ meson formation time corresponding to the $\gamma\to \rho$ 
transition is $\tau\sim 2E_\gamma/m^2_\rho$ 
in the target $A$ rest frame, and is therefore very large (see ~\cite{Gr,Io}). For Pb-Pb heavy ion collisions at an LHC energy $\sqrt{s_{nn}} \sim 2.76 $ TeV corresponding to 7 TeV LHC energy, then $E_\gamma\simeq 1.2 $ TeV for $Y_\rho=0$. That is 
\be
\tau~\simeq~4\times 10^3 ~\GeV^{-1} \simeq ~~10^3~{\rm fm} ~\gg ~ R_A~\simeq ~7~{\rm fm}.
\label{tau}
\ee
 So the size  of the target in negligible in comparison with the $\rho$ formation time. We sketch the situation in Fig.~{\ref{fig:2}}.

Since the size of the target ion $A$ is negligible in comparison to the $\rho$ formation time, we may neglect the possibility of the $\gamma \to \rho$ transition {\it occurring within the target}.  It would be a small addition to the coherent production rate.

Therefore the differential cross section $d\sigma_{\rm el} (AA\to A\rho A)/dt$ 
  will reveal a distinctive diffractive structure (with a sequence of dips, just as in optics). Such a dip structure was first observed by the STAR Collaboration \cite{Debbe:2012aa,Klein:2018grn,Adamczyk:2017wyc} and was also seen in \cite{ALICE} .We have explained why the incoherent contribution to heavy ion scattering should be small in comparison to  coherent production
 and consider corrections to the description of the exclusive process $AA\to A+\rho +A$ illustrated in Fig.~\ref{fig:1}. 
First, the interference \cite{KN2} and, secondly, the incoherent processes which are hard to completely exclude from the experimentally observed cross section. 
 What do we mean by interference? Besides Fig.~\ref{fig:1} there is a second exclusive diagram in which $A_{\rm target}$ becomes the photon emitter. The exclusive cross section therefore contains an interference term between  these two contributions. 
 This two-source interference was observed by the STAR Collaboration  in $\rho$ photoproduction in gold-gold collisions \cite{Abelev:2008ew}.

\section{Vector Dominance Model }

The photon emitted from the `beam' ion transforms into hadronic states in two stages. First, the $\gamma$ creates a point-like $q\bar{q}$ pair, which then after some time forms the hadronic system. Using the Vector Dominance Model \cite{Sakurai:1960ju,Bauer:1977iq} this system is described by the sum of $\rho , ~\omega, ~\phi ,~J/\psi~...$ vector meson resonances.  The model assumes that in the low-mass region the first few resonances saturate the amplitude.  Then the $\gamma\to V$ transition vertex is calculated from the known $e^+ e^-$ decay width, $\Gamma^V_{ee}$,
of the corresponding resonance
\be
\gamma^2_V~=~\frac{3\Gamma^V_{ee}}{\alpha^{\rm QED} M_V};\quad\;\;\;\;\;\gamma^2_V~\simeq~3.8\cdot 10^{-3}\;\;\;\mbox{for the $\rho$ meson}.
\ee
As a result, the cross section for exclusive $V$ meson photoproduction may be written
\be
\sigma(\gamma p\to Vp)~=~\gamma^2_V~\sigma_{\rm el}(Vp\to Vp).
\ee
Thus we may use the HERA data to determine $ \sigma_{\rm el}(Vp\to Vp)$. The experiments found that the cross section increases with the centre-of-mass energy, $W$. In particular for $\rho$ production we have \cite{Wrho}
\be
\sigma(\gamma p\to \rho p)~=~\sigma_0~(W/W_0)^{0.22}
\label{eq:W}
\ee
where $W_0=200$ GeV and $\sigma_0=15~\mu$b. In terms of the total cross section, this corresponds to $\sigma_{\rm tot}(\rho p)=27.5$ mb at $W$=200 GeV.  We use this result in our numerical estimates.

Note that the  cross section extracted from photoproduction ($\gamma p\to \rho p$) data is a bit lower than the true $\rho p$ cross section since it includes configurations where the $q\bar q$ pair was created relatively close to the target and the full $\rho$-meson wave function has insufficient  time to form completely. On the other hand this is just the value one has to apply in such a calculations.
 The fact that the heavy ion thickness (depth) of the order of 10 fm is larger than that for the proton does not change the situation. This difference is negligible in comparison with the formation time $\tau\sim 1000$ fm (see eq.(\ref{tau})).

\section{Photon flux}
The photon flux emitted by the heavy `beam' ion $A$ can be expressed in momentum space as
\be
\frac{dN_\gamma}{dx}~=~\frac{Z^2 \alpha_{\rm QED}}{\pi x} \int dk^2_t\frac{k^2_t F_A^2(k^2_t)}{{(k^2_t+(xm_n)^2})^2}
\label{eq:flux}
\ee
where $Z$ and $F_A$ are respectively the charge and the form factor of the heavy ion $A$, and $x$ is the fraction of the {\it nucleon} energy carried by the photon; $m_n$ is the mass of the nucleon. Since we are working in the very low $x$ region we have neglected terms proportional to higher powers of $x$ in (\ref{eq:flux}).  Indeed for energies $\sqrt{s_{nn}}=2.76$ TeV at the LHC we have
\be
x~=~\frac{m_\rho}{\sqrt{s_{nn}}}~\simeq ~0.3\times 10^{-3}
\ee
for the central ($Y_\rho = 0$) 
production of a $\rho$ meson. Actually the integral in (\ref{eq:flux}) runs logarithmically as $\int dk_t^2/k_t^2$  from 
$k_t\sim xm_n\sim0.3$ MeV 
up to $k_t\sim 1/R_A\simeq30$ MeV, the latter value is limited by the form factor $F_A$.  To be very precise we should note that the flux depends on the particular position of the photon in impact parameter $b_\gamma$-space with respect to the centre of the parent ion.
Outside the spherical ion ($b_\gamma>R_A$) the flux takes the form
\be
\frac{d^3N_\gamma}{dxd^2b_\gamma} ~=~ \frac{Z^2 \alpha^{\rm QED}}{x\pi^2 b_\gamma^2}~ (xm_n b_\gamma)^2~K_1^2(xm_n b_\gamma),
\label{eq:fluxb}
\ee
where $K_1(z)$ is the modified Bessel function. Note that $K_1(z)\to1/z$ as $z\to 0$, thus the last product in (\ref{eq:fluxb}) approaches a constant. 

The $b_\gamma$ representation is convenient to account for the survival factor of the rapidity gaps. At fixed $b_\gamma$ the survival factor $S^2~=~$exp$(-\Omega_{AA})$. Usually this factor is replaced by $\theta(b_\gamma-2R_A)$. However, an explicit calculation \cite{chic3} shows that actually the value of $S^2$ is still very small
 even for a bit larger $b_\gamma$.  For Pb-Pb heavy ion collisions it can be approximated by $\theta(b_\gamma-17$ fm). Thus the full photon flux will be given by the integration of (\ref{eq:fluxb}) over the region of $b_\gamma$ larger than 17 fm. To about 10$\%$ accuracy it may be written as
\be
\frac{dN_\gamma}{dx} ~=~ \frac{Z^2\alpha^{\rm QED}}{x\pi}~{\rm ln}\left(\frac{1}{(4R_Axm_n)^2}\right),
\ee
where $R_A\simeq 7$ fm.

\section{The $\rho A$ interaction}
The elastic $\rho A$ amplitude can be written in the Glauber eikonal approximation\footnote{We do not include the inelastic Glauber corrections since the
effect of inelastic shadowing is
almost compensated by the effect of short-range correlations in the wave
function of the target
nucleus~\cite{Frankfurt:2015cwa, CiofidegliAtti:2011fh}.} as
\be
A_{\rho A}(b)~=~i(1-e^{-\Omega(b)/2})
\label{eq:elamp}
\ee
where here $b$ is the impact parameter of the $\rho$ meson with respect to the heavy ion and the opacity $\Omega(b)$ is given by
\be
\Omega(b)=T_A(b)\sigma_{\rho n}~\eta~~~~~{\rm with}~~~~~T_A(b)=\int_{-\infty}^\infty dz \rho_N(z,b).
\label{eq:Omega}
\ee
Here $\sigma_{\rho n}$ is the total cross section of the $\rho$-nucleon interaction parametrized according to 
(\ref{eq:W}) 
 and
\be
\eta~=~1-i{\rm tan}(\pi\Delta/2)
\ee 
is the signature factor which accounts for the phase of the even-signature (Pomeron) amplitude\footnote{ Even-signature means that the amplitude is symmetric under the permutation of $s$ to $u$ (see, for example, \cite{PDBC}).} which increases with $s$ as $s^{1+\Delta}$. For the parametrization (\ref{eq:W}) we have $\Delta=0.22/4$. 

In general, there may be excitations of the intermediate states. This effect can be accounted for using the Good-Walker formalism \cite{GW}. However, we neglect this relatively small effect in the present paper.

For the nucleon density distribution, $\rho_N$, in the heavy ion we use
the Woods--Saxon form~\cite{Woods}
\be
\rho_N(r)= \frac{\rho_0}{1+\exp{((r-R)/d)}}\;,
\label{eq:sim}
\ee
where the parameters  $d$ and $R$ respectively characterise the skin thickness and the radius of the nucleon density in the heavy ion. For $^{208}$Pb
we take the recent results of~\cite{Tarbert,Jones}
\begin{align}\nonumber
R_p &= 6.680\, {\rm fm}\;, &d_p &= 0.447 \, {\rm fm}\;,\\ \label{eq:pbpar}
R_n &= (6.67\pm 0.03)\, {\rm fm}\;, &d_n &= (0.55 \pm 0.01) \, {\rm fm}\;.
\end{align}
The nucleon densities, $\rho$, are normalized to 
\be
 \int\rho_p(r)d^3r=Z \;, \qquad \int\rho_n(r)d^3r=N\;,
\ee
for which the corresponding proton (neutron) densities are $\rho_0 = 0.063$ (0.093) ${\rm fm}^{-3}$.

Since the optical density is quite large, the scattering amplitude has a black disc form
\be
A_{\rho A}(b)=i\theta(R_A-b)
\label{eq:sim}
\ee
up to the edge region, $b=R_A\pm d$. After we take the Fourier transform
\be
A_{\rho A}(p_t)~=~ 2s\int d^2b~ e^{i\vec{b}\cdot \vec{p_t}}~A_{\rho A}(b)
\label{eq:amp2}
\ee
we obtain the $\rho A$ differential cross section
\be
\frac{d\sigma_{\rho A}}{dp_t^2}=\frac{|A_{\rho A}(p_t)|^2}{16\pi s^2},
\label{eq:X}
 \ee
which reveals a diffractive dip structure analogous to that observed in optics from light scattering by a black disc. 

Recall that actually we use the complete $\rho A$ amplitude (\ref{eq:elamp},\ref{eq:Omega}) and not the simplified form (\ref{eq:sim}). This is the conventional Glauber eikonal approach
which in the EIC review (entitled Electron Ion Collider: the Next QCD Frontier))~\cite{EIC} and in Sartre event generator~\cite{sartre}
was called `saturated model'.\footnote{This terminology is confusing. In fact the so-called `non-saturated' model' of \cite{sartre} corresponds to the conventional impulse approximation which is
well known not to be applicable for heavy ion interactions.}

The result corresponding to $\rho$Pb$\to \rho$Pb scattering is shown by the lowest (blue) curve in Fig.~\ref{fig:11}. This figure is for the purely coherent contribution to $\rho$ production and will be discussed in Subsection~\ref{sec:6.1}.  Indeed, in Section~\ref{sec:6} we show detailed plots of the predictions for dip structure before and after including the incoherent component.
\begin{figure} [h]
\begin{center}
\includegraphics[trim=0cm -0cm 0cm 10cm,scale=0.7]{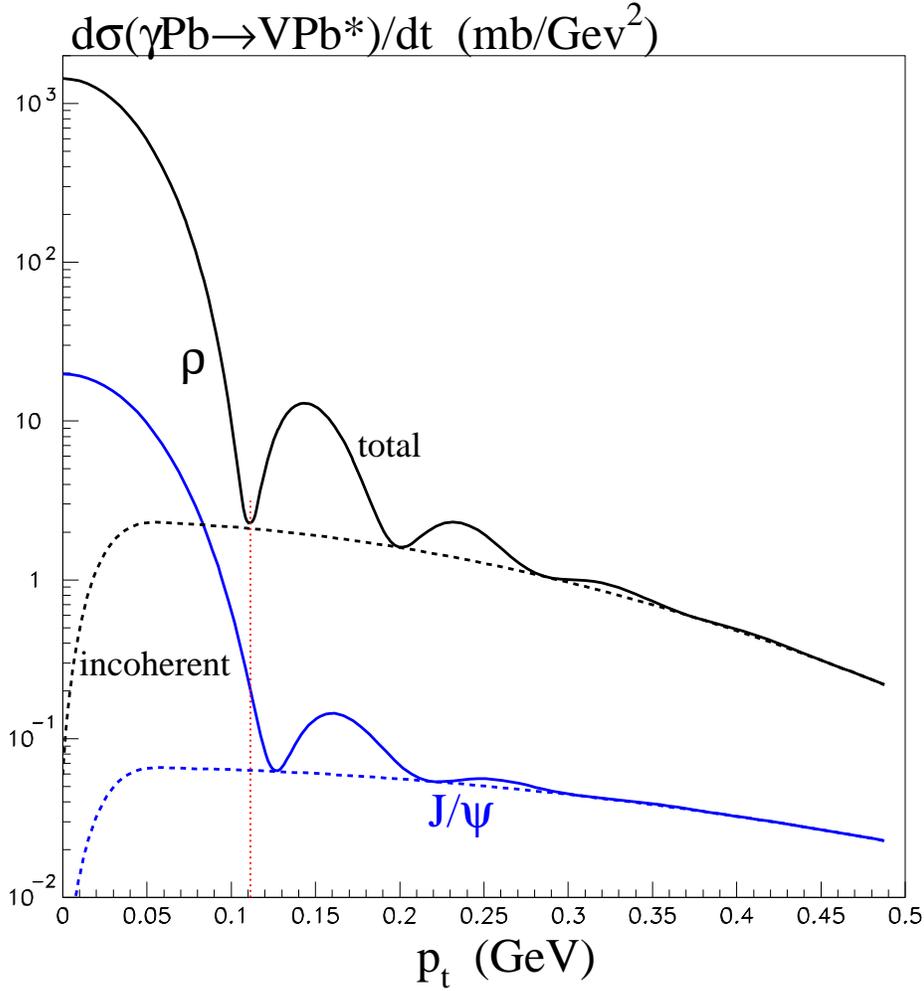}
\caption{\sf Just for illustration we show the simple prediction of the differential cross section before we include the interference effects 
in the coherent contribution and before we give a detailed study of the incoherence effects. 
 Here we also show the results for $J/\psi$ production in heavy ion collisions. Since the $J/\psi p$ cross section is smaller the dip positions  move to a larger $p_t$.}
\label{fig:3}
\vspace*{-0cm}
\end{center}
\end{figure}

It is informative first to show in Fig.~\ref{fig:3} the differential cross section $d\sigma /dt$ for $\gamma $Pb$ \to V$Pb$^*$ for both $V=\rho$ and $J/\psi$ production.  Interference effects are not yet included in the coherent contribution and a very simple estimate is made of the incoherent component. In this idealized case the dip structures are clearly evident.

\section{Beyond the leading contribution}
To obtain the result for the full process $AA\to A+\rho +A$ we have to multiply (\ref{eq:X}) by the photon flux (which already accounts for the survival factor $S^2=\theta(b-17$ fm)) and for the probability of the $\gamma \to \rho$ transition. However, to be precise we have to account for a few additional effects.

\begin{figure} [h]
\begin{center}
\includegraphics[trim=-2cm 8cm 0cm 0cm,scale=0.5]{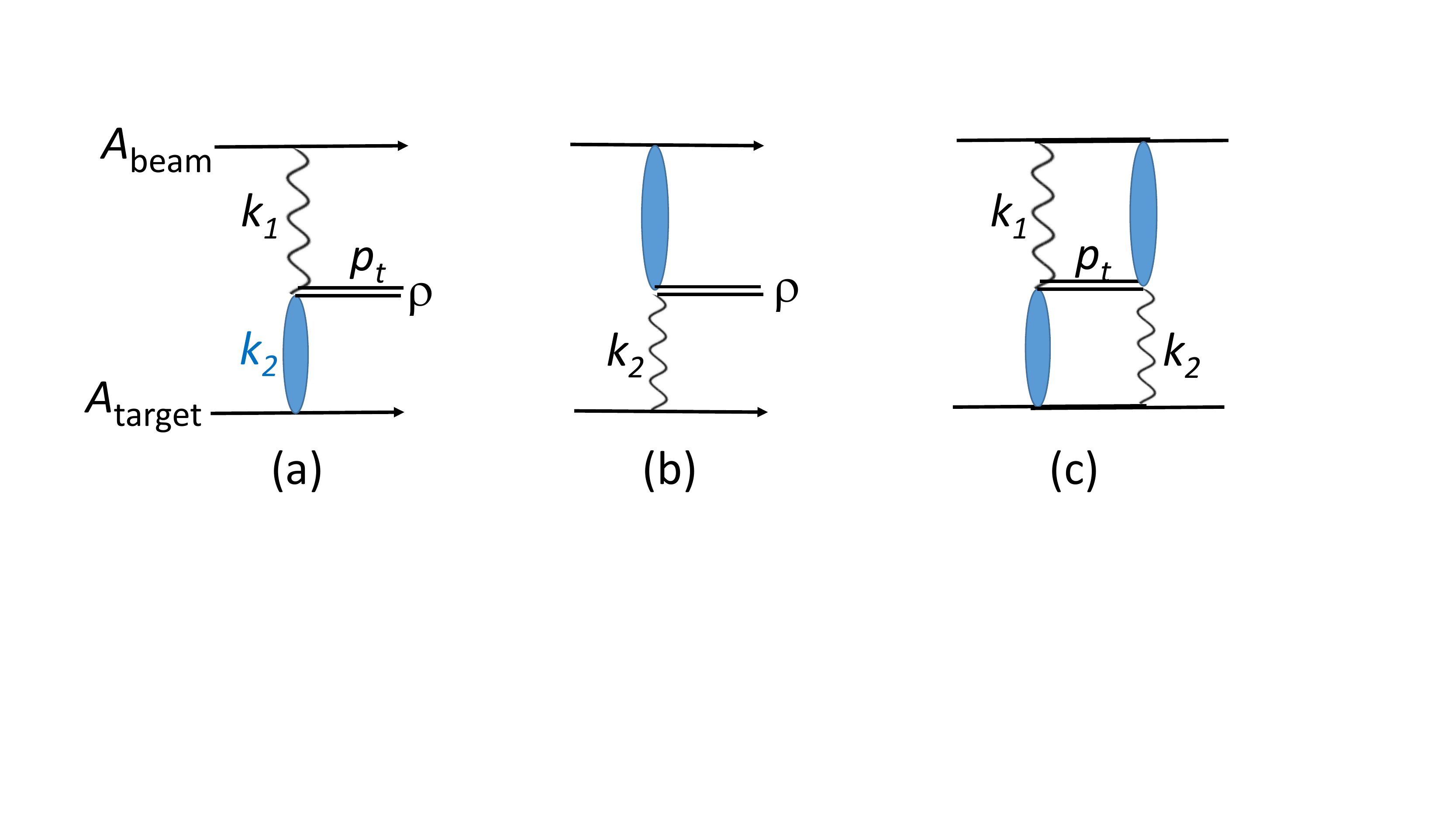}
\caption{\sf There are two diagrams, (a) and (b), contributing to the process $AA\to A+\rho +A$. Diagram (c) shows the interference of the amplitudes corresponding to these diagrams in which the photon $(k_1,k_2)$ is emitted from the $A_{\rm beam}$ or $A_{\rm target}$ respectively. Note that while diagrams (a) and (b) correspond to amplitudes, diagram (c) shows the interference contribution to the cross section. }
\label{fig:4}
\vspace*{-0cm}
\end{center}
\end{figure}
\subsection{Interference}
For heavy ion $AA$ collisions there are two diagrams, Figs.~\ref{fig:4}(a,b) to consider, arising from the photon emitted by either $A_{\rm beam}$ or by $A_{\rm target}$.  In general there will be interference between the
two corresponding amplitudes shown in diagram (c).  At first sight it looks as  the interference is very small, since the $k_t$ integral (\ref{eq:flux}) for diagram (c)
\be
\int dk_{t1}^2~\frac{{\vec k}_{t1}\cdot {\vec k}_{t2}}{(k_{t1}^2+(x_1m_n)^2)~(k_{t2}^2+(x_2m_n)^2)}
\label{eq:int}
\ee
loses its logarithmic form when 
${\vec k}_{t2}=-{\vec p}_t-{\vec k}_{t1}\neq {\vec k}_{t1}$. However, for very small $p_t$ ($p_t\ll k_{ti}$) the logarithmic structure of the integral (\ref{eq:int}) is restored. That is, we obtain a logarithm by integrating over $k_t$ from $p_t$ up to $1/R_A$. This should be compared with the case of diagrams (a),(b) where the logarithm comes from the $k_t$ interval from $x_im_n$ to $1/R_A$. Moreover, due to the negative P-parity of $\rho$ the interference term is destructive. Therefore in the symmetric case (where the rapidity of the meson $Y_\rho =0$) the cross section vanishes at $p_t=0$.

In the leading log approximation we can see the effect of the interference diagram (c) in the expression below, which shows the sum of the contributions of the three diagrams which arise in the photon flux (\ref{eq:flux})
\be
2~[~{\rm ln}(2R_A x_1 m_n)+{\rm ln}(2R_A x_2m_n)~]~-~ 2~{\rm ln}(2R_A^2 (x_1^2+x_2^2) m_n^2 +p_t^2)
\ee
The first term in [...] arises in the sum of diagrams (a),(b), while the latter term arises from diagram (c). The interference  effect was first considered in \cite{KN2} and was  confirmed by the STAR \cite{Abelev:2008ew}
 experiment. Actually the interference 
 is only visible at very small $p_t$ in the symmetric configuration ($Y_\rho \simeq 0)$.

\subsection{Dependence on photon impact parameter $b_\gamma $}

Recall that the photon flux ({\ref{eq:fluxb}) has a dependence on the impact parameter $b_\gamma$ of the photon. That is for different values of $b_\gamma$ 
 the amplitudes (\ref{eq:elamp}) and (\ref{eq:amp2}) 
 should be multiplied by a slightly different photon fluxes. An explicit calculation in $b_\gamma,~b$ space shows that this slightly deforms the shape of the diffractive peak.  Nevertheless the effect is quite small, see 
 Figs.~\ref{fig:11},\ref{fig:12}.

Besides this, strictly speaking, we have to account for the fact that the observed transverse momnetum $p_t$ of $\rho$ meson is not exactly equal to the
momentum ($k_{2t}$ in the case of the configuration shown in Fig.~\ref{fig:4}(a) transferred in $\rho A$ collision amplitude. It is slightly washed out by the momentum of photon. However this effect is very small as well.

\subsection{Incoherent contribution}
Much more important is the contribution from incoherent processes where the heavy ions (or even nucleons in the heavy ions) break up. Formally this contribution can be excluded by rejecting such events by observing the decay products. Unfortunately in the present experiments this would be challenging. Therefore the diffractive picture of the coherent $\rho A$ differential cross section (with its dips and peaks) will sit on top of an incoherent background. As mentioned in the Introduction, the cross section of ion dissociation, $\sigma_{\rm diss}\simeq 2\pi R_A d$ is suppressed in comparison with the elastic cross section, $\sigma_{\rm el}\simeq \pi R_A^2$, by the small width, $d$ of the peripheral ring. The suppression factor is 
$2d/R_A \simeq 1/6$. However these incoherent cross sections, especially in the case of nucleon dissociation, have a very flat $p_t$ dependence. 
Therefore already in the region of the second and third coherent peaks they tend to obscure the diffractive peak structure.

The incoherent cross section can be calculated as (see~\cite{chic3} for more details)
\begin{equation}
\label{incoh}
\frac{d\sigma^{\rho A}_{\rm incoh}}{dp^2_t}~=~\int d^2b ~T_A(b)\frac{d\sigma^{\rho n}}{dp^2_t}
\exp(-\Omega(b))~[1-F^2_A(p^2_t)]\ ,
\end{equation}
where the $\rho A$ collision opacity $\Omega(b)$ is given by (\ref{eq:Omega}) and 
$d\sigma^{\rho n}/dp^2_t$ is the elementary $\rho$-nucleon cross section, for which we use the same parametrization (\ref{eq:W}) of the HERA data, with a $t$-slope $B=10$ GeV$^{-2}$~\cite{Wrho}. The last term in square brackets accounts for the fact that for a very low $p_t$ there is some probability (given by the ion form factor, $F_A$, squared) not to destroy the ion.

In the case of nucleon $n\to n^*$ dissociation we allow for the excitation of nucleon resonances and for relatively high mass $(n\to M_X)$ dissociation. However the value of $M_X$ should not be so large that the particle produced by $M_X$ hadronization fills the rapidity gap (that is, can be observed in the detector).  We assume that the total cross section of dissociation is about the same as that for elastic  scattering 
as was measured by HERA (see the discussion given in~\cite{HERA-diss}) for which we take the slope $B_{\rm diss}=3$ GeV$^{-2}$. (Of course in the case of $n\to n^*$ dissociation the last $[1-F^2_A]$ factor must be omitted.)
\begin{figure} [h] 
\begin{center}
\includegraphics[trim=0cm -0cm 0cm 9cm,scale=0.6]{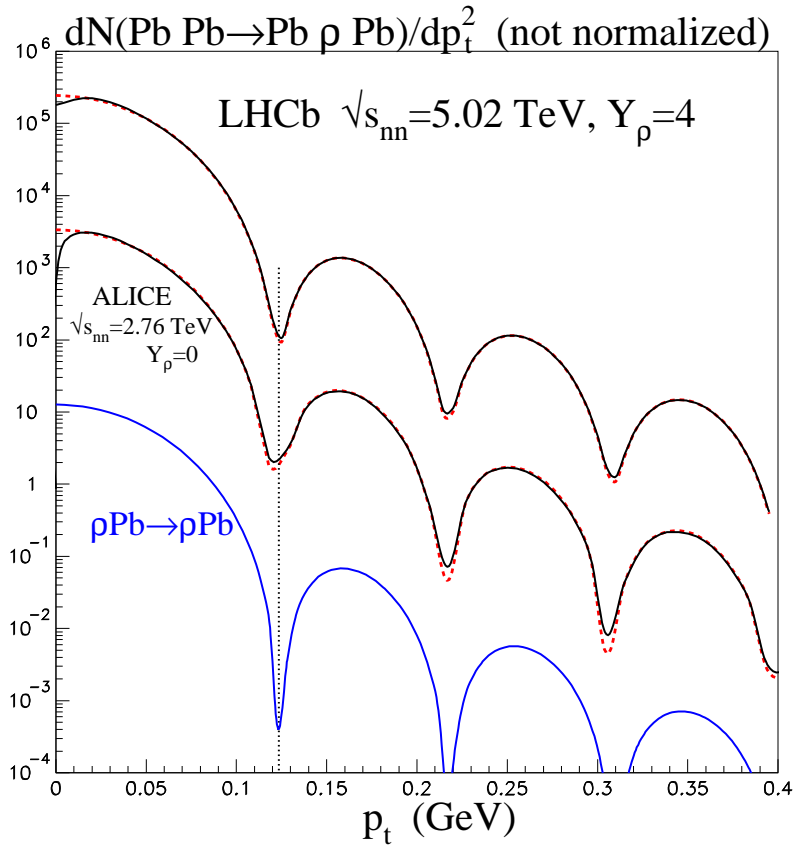}
\caption{\sf The differential cross section of processes Pb Pb $\to$ Pb $\rho$ Pb  and $\rho$  Pb$\to\rho$  Pb. The whole amplitude was calculated in impact parameter ($b$) representation accounting for interference and survival effects, and not assuming factorization. For the top two curves the red dashed curves show the prediction before interference effects are included. We see that the effect of interference is tiny; it only affects very small $p_t$, or in the symmetric ($Y_\rho=0$) case it fills in the dips a little. The vertical line is drawn to better observe the shift of the first dip.}
\label{fig:11}
\vspace*{-0cm}
\end{center}
\end{figure}

\section{Predictions relevant to experimental observations  \label{sec:6}}

In the first subsection we discuss (as illustrated by Figs.~\ref{fig:11} and \ref{fig:12}) the detailed properties of the dip structure of the {\it purely coherent} contribution of $\rho$ production in Pb-Pb collisions. Then in the following subsection we include the incoherent contribution and show in Fig.~\ref{fig:13} how it could mask the observation of the second and third and higher diffractive dips.

\subsection{The dip structure of the coherent contribution  \label{sec:6.1}}
The results of the explicit computation of pure coherent $\rho$ production in Pb-Pb high-energy collisions made in $b$-representation are shown in Figs.~\ref{fig:11} and \ref{fig:12}. Recall that the calculations were performed in the impact parameter ($b$) representation which facilitate inclusion of interference and survival effects. 

We denote the impact parameter of the $\rho$ meson with respect to the beam ion by $b_1$ and
with respect to the target by $b_2$. Then the amplitude reads
\begin{equation}
\label{b-space}
A(k_1,k_2)~=~\int d^2 b_1 d^2 b_2 A(b_1,b_2)\exp(i\vec b_1\cdot\vec k_1)\exp(i\vec b_2\cdot\vec k_2)\ ,
\end{equation} 
where $k_1$ and $k_2$ are the transverse momenta of the beam and the target ions
respectively. Thus in this way in the computation we do not neglect the value of photon transverse momentum ($k_1$ or $k_2$).

In Fig.~\ref{fig:11} the sum of the contributions of diagrams (a) and (b) of Fig.~\ref{fig:4} is shown by the red dashed curves, while the black curves show the result after including  the interference contribution corresponding to diagram (c). As expected, the effect of interference is only visible at very small $p_t$. For comparison the $p_t$ distribution of elastic $\rho$ Pb scattering, which plays the role of the subprocess is shown by the lowest (blue) curve at the energy corresponding for $Y_\rho=4$ to the largest contribution configuration (say, Fig.~\ref{fig:4}(a)).
\begin{figure} [h]
\begin{center}
\includegraphics[trim=0cm -0cm 0cm 9cm,scale=0.6]{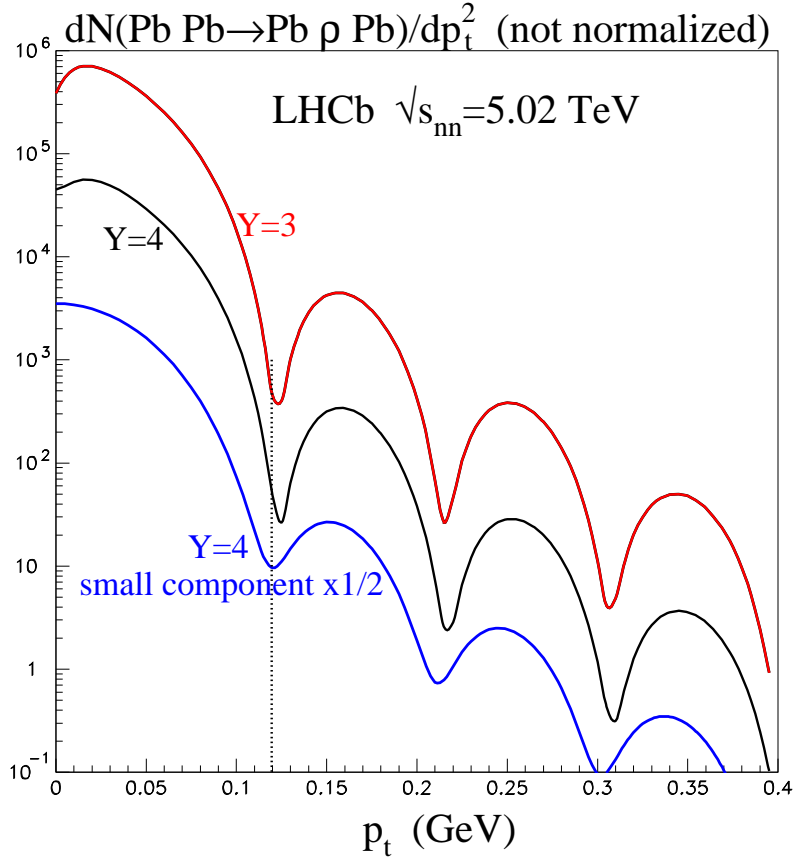}
\caption{\sf The differential cross section of the process Pb Pb $\to$ Pb $\rho$ Pb for kinematics accessible to the LHCb detector. The whole amplitude was calculated in impact parameter representation accounting for interference and survival effects, and not assuming factorization. The small component corresponds $|A^{(b)}|^2$ in the configuration when the $\rho$ meson goes in the direction of $A_{\rm target}$; if $Y_\rho=-4$ then the curve would correspond to $|A^{(a)}|^2$; here (a) and (b) refer to diagrams (a) and (b) in Fig.~\ref{fig:4}. The vertical line is drawn to better observe the shift of the first dip. None of these curves are normalised and are simply to show
the $p_t$  behaviour.    However, the factor 0.5 included on the lowest  `small component'
curve is to keep  its  normalisation  the  same  as  that  for  the  `total’
$Y_\rho=4$ curve shown here;
that is, to show the relative size of the `large' and `small' components.}
\label{fig:12}
\vspace*{-0cm}
\end{center}
\end{figure}
It is seen that the $p_t$ distribution is very similar for the reactions  Pb Pb $\to$ Pb$+\rho +$Pb and the $\rho$ Pb $\to\rho$ Pb. In particular, the positions of the dips are exactly the same. The interference washes out the dips a little, while the $b_\gamma$ dependence of photon flux only deforms the peaks by a very small amount.

At a larger subprocess energy  the dip position moves to a bit smaller $p_t$.
This is in analogy with the shrikage of the diffractive cone. In our case the shift of the dip position reflects the fact that at a larger energy we have a larger 
$\rho n$ cross section and therefore the $\rho$ meson feels the edge of target at a bit larger value of $b$. In other words, the `effective' size of the disc increases and correspondingly the value of $p_{t,{\rm dip}}\propto 1/R$ decreases.
 
 In Fig.~\ref{fig:12} we compare the distributions at two rapidities ($Y_\rho=3$ and $Y_\rho=4$) corresponding to kinematics accessible to the LHCb experiment. At $Y_\rho=4$ the largest contribution comes from the component with a smaller $\rho$-Pb energy 
 \be
 s(\rho {\rm Pb})\propto \exp({-Y_\rho})
 \ee
  and, as expected we observe the dip at a larger $p_t$ than for $Y_\rho=3$. 
  On the other hand the component with a larger $\rho$ P sub-energy, which gives a smaller contribution to the total cross section 
   due to a smaller photon flux, has a dip at a smaller value of $p_t$.

\begin{figure} [h]
\begin{center}
\includegraphics[trim=0cm -0cm 0cm 9cm,scale=0.65]{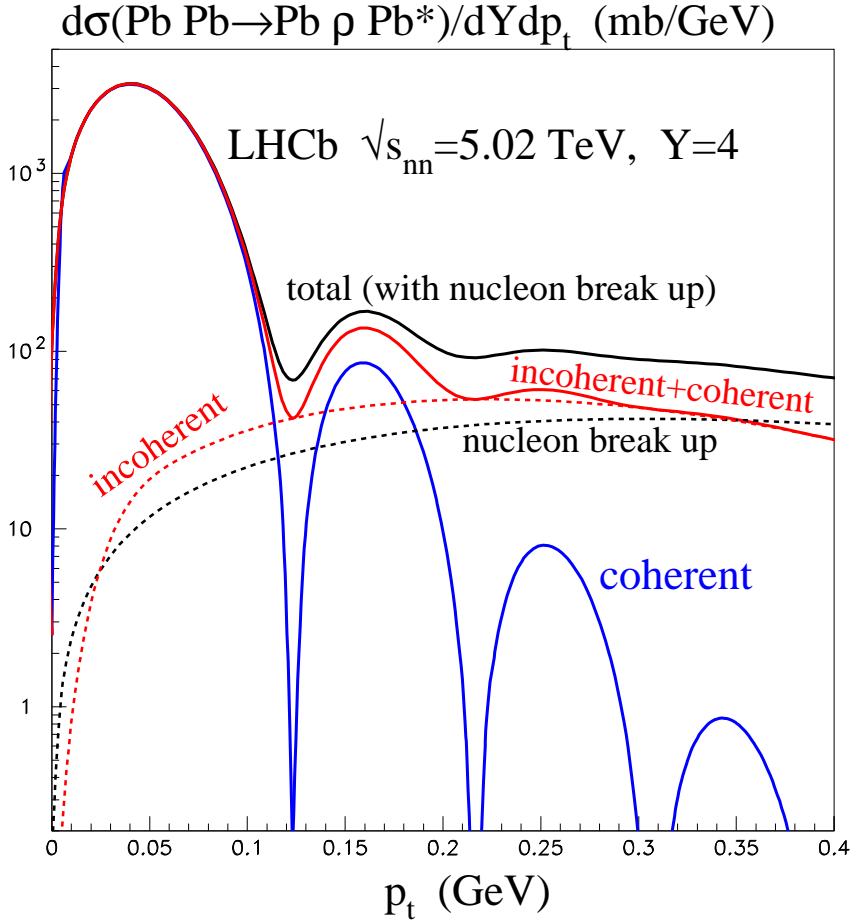}
\caption{\sf The upper curve shows the differential cross section $d\sigma/dY_\rho dp^2_t$ for the process Pb Pb $\to$ Pb $\rho$ Pb$^*$ at $Y_\rho=4$ and $\sqrt s_{nn}=5.02$ TeV.  It is the sum of the coherent and the incoherent components. The part of the incoherent component due to nucleon break up is indicated; the remaining incoherent component is due the dissociation of the heavy ion (denoted Pb$^*$).}
\label{fig:13}
\vspace*{-0cm}
\end{center}
\end{figure}
\subsection{Including the incoherent component} 
 Finally, the upper curve in Fig.\ref{fig:13} shows the prediction for $d\sigma/dY_\rho dp^2_t$ for the process Pb Pb $\to$ Pb $\rho$ Pb$^*$.
Note that here we do not plot $dN/dp^2_t$ but instead the $d\sigma/dp_t$ distribution,
which is now correctly normalized.  The dashed curves below the upper curve indicate the role of the incoherent components.
Of course for $p_t>200$ MeV the large incoherent cross section strongly  masks
the diffractive dips. We see that the dip structure is a little more evident when we have possibility to reject events where the nucleon was broken up and produced some additional particles in a forward (large rapidity) region. Then we replace the black curve by the lower (red) curve where the third maximum is quite visible. Having a very good detector in the large rapidity region, the LHCb Collaboration has a chance to reject also part of the events 
caused by the {\it elastic} $\rho+n$ scattering subprocess, which due to a large value of  $p_t$ (in comparison with $1/R_A$) still breaks up the target ion. Then the coherent component with a series of diffractive dips will be observed even better.

Recall that we need to reject extra secondaries only in the region of the target dissociation, that is in a forward rapidity region (for the LHCb case) where the dominant contribution comes from the interaction of vector meson with the ion going in the forward ($\rho$) direction (in the laboratory frame). The probability of dissociation of the other ion (which emits the photon) is suppressed by a factor of $1/Z$, and thus should be very small.

\section{Further effects}
For completeness, we mention a few points which were not implemented in the present calculation.

\subsection{Direct photoproduction of $\pi^+\pi^-$ pair}
To be precise we have to consider also the direct production of a $\pi^+\pi^-$ pair
 (directly arising from $\gamma\to\pi^+\pi^-$) followed by interaction of the pair with target ion.
 The main contribution comes from the interference of the `direct $\pi\pi$' amplitude with the real part of the Breit-Wigner $\rho$ meson term~\cite{Soding}.
 This interference enhances the cross section at low mass $M_{\pi\pi}<M_\rho$ but is destructive for  $M_{\pi\pi}>M_\rho$. Since the $\pi\pi$ pair has its own size the absorption of at least one pion can take  place at a larger impact parameter $b$ than that for the absorption of the $\rho$ meson. Correspondingly, for `direct' $\pi\pi$ production the position of  first (and the next) diffractive dip(s) should be observed at a lower $p_t$. Therefore we expect a bit smaller value of $p_{t,{\rm dip}}$ for low $M_{\pi\pi}<M_\rho$, a bit larger value 
of $p_{t,{\rm dip}}$ for  $M_{\pi\pi}>M_\rho$ and again a smaller $p_{t,{\rm dip}}$ at $M_{\pi\pi}>0.85-0.9$ GeV where the `direct' amplitude starts to dominate.

 \subsection{The production of two vector mesons}
Since the photon flux radiated by a lead ion is rather large and the cross section of vector meson production is huge there should be a noticeable probability to observe events where two vector mesons (say, $\rho\rho$ or $\rho\phi$) are produced simultaneously. Such a possibility was discussed in ~\cite{KN,Klein2,Baur:2003ar}.  Recall that the main contribution for ultraperipheral processes comes from the region of very large $b_\gamma \gg b$. On the other hand the vector meson polarization vector is directed along $\vec b_\gamma$. In the case of two vector meson production on the same target the separation between
the position of each meson $|\vec b_{1\gamma}-\vec b_{2\gamma}|<2R_A$. That is two polarization vectors are almost parallel. Experimentally we cannot measure the impact parameter $b_\gamma$ but it should be possible to observe the corresponding correlation in the decays of the two vector mesons.

\subsection{Giant dipole resonance}
As was emphasized in~\cite{Baur:2003ar,BKN} for such a large value of $Z=82$ there is a  probability for the excitation of a Giant dipole resonance (GDR) due to multiphoton  Coulomb exchange between the two lead ions. This probability depends on the ion-ion  separation  in impact parameter $b$ space and for the case of $\rho$ meson production at the LHC energies it can reach 7 - 10 \% for each ion.
 The GDR decays emitting a neutron. This  was observed and confirmed in ALICE experiment~\cite{ALICE} where  the fraction of events without an additional  neutrons detected in the Zero Degree Calorimeter was about 85\%.

\section*{Acknowledgements}
 We thank Ronan McNulty and Tara Shears for stimulating our interest in this problem and for useful discussions, and Spencer Klein for clarifying comments.
 MGR thank the IPPP at the University of Durham for hospitality.
VAK acknowledges support from a Royal Society of Edinburgh Auber award.

\enddocument